\documentclass[showpacs,prl,twocolumn,aps,superscriptaddress,preprintnumbers,letterpaper]{revtex4}
\usepackage{amsmath,amssymb}
\usepackage{epsfig}
\usepackage{graphicx}
\usepackage{amsmath}
\usepackage{amsfonts}
\usepackage{epstopdf}
\usepackage{color}

\newcommand{\be}{\begin{equation}}
\newcommand{\ee}{\end{equation}}
\newcommand{\bear}{\begin{eqnarray}}
\newcommand{\eear}{\end{eqnarray}}

\newcommand{\nn}{\nonumber\\}
\newcommand{\ba}{\begin{array}}
\newcommand{\ea}{\end{array}}

\begin{document}

\title{Chiral and Gravitational Anomalies  on Fermi Surfaces}

\author{G\"ok\c ce Ba\c sar}
\email{basar@tonic.physics.sunysb.edu} 
\affiliation{Department of Physics and Astronomy, Stony Brook University, Stony Brook, New York 11794-3800, USA}
\author{Dmitri E. Kharzeev}
\email{dmitri.kharzeev@stonybrook.edu}
\affiliation{Department of Physics and Astronomy, Stony Brook University, Stony Brook, New York 11794-3800, USA}
\affiliation{Department of Physics, Brookhaven National Laboratory, Upton, New York 11973-5000, USA}
\author{Ismail Zahed}
\email{zahed@tonic.physics.sunysb.edu}
\affiliation{Department of Physics and Astronomy, Stony Brook University, Stony Brook, New York 11794-3800, USA}

\date{\today}

\begin{abstract}
A Fermi surface threaded by a Berry phase can be described by the Wess-Zumino-Witten (WZW) term. After gauging, it produces a five-dimensional Chern-Simons term in the action. We show how this Chern-Simons term captures the essence of the Abelian, non-Abelian, and mixed gravitational anomalies in describing both in- and off-equilibrium phenomena. In particular we derive a novel contribution to the Chiral Vortical Effect that arises when a temperature gradient is present. We also discuss the issue of universality of the anomalous currents. 
\\
\\
\centerline{{\it To the memory of Gerry Brown}}
\end{abstract}


\maketitle

\setcounter{footnote}{0}

\vskip0.2cm

{\bf 1.\,\,Introduction.}
Chiral anomalies have a topological origin that is concisely captured by the Wess-Zumino-Witten (WZW) terms 
\cite{WESS,WITTEN}.  They are at the origin of the anomalous decays of neutral pions and most of the 
vector meson decays caused by the parity odd contributions to the chiral currents~\cite{WESS,Zahed:1986qz}.
Because of their topological nature, these anomalies are not renormalized by interactions or the presence of 
matter~\cite{BARDEEN}. Chiral flavor anomalies in dense QCD were noted initially in~\cite{CFL}, 
and more recently  in holography~\cite{ADS,ANTON}.  They can be described by gauged WZW effective 
actions~\cite{MISHA,SSZ,NAIR} and anomalous hydrodynamics~\cite{SON1,Kharzeev:2011ds}.
 The
chiral and mixed gravitational anomalies in condensed matter physics with applications to superfluid $^3$He-A were  
studied by Volovik and others, see~\cite{VOLOVIK} and references therein, and in plasmas in \cite{GRAVITY,Landsteiner:2011iq}. Triangle anomalies affect hydrodynamics \cite{HYDRO}, Fermi surfaces with Weyl points ~\cite{VOLOVIK,SON,WEYL,Basar:2013iaa,Son:2013rqa}, and possibly graphene ~\cite{GRAPHENE}. 
The Fermi surfaces with Weyl points possess the Berry phase leading to the chiral anomaly. 
Indeed, in the presence of Berry phase the chiral currents develop gauge invariant anomalous contributions
\cite{VOLOVIK,SON}. These contributions can be derived from  the WZW 
terms that in turn can be cast in the form of a five-dimensional Chern-Simons (CS) term~\cite{VOLOVIK,ZAHED}.  

In this letter we extend the 5D CS approach to describe the in- and out-of-equilibrium dynamics of systems with chiral Fermi surfaces. Our original results consist
of:  i) a new contribution to the Chiral Vortical Effect (CVE) arising when a temperature gradient is present; ii) new temperature-induced effects in 
non-Abelian chiral superfluids; 
iii) a new semi-classical method of introducing the non-Abelian and mixed gravitational anomalies through the monopole in momentum space; 
iv) a complete treatment of anomalous off-equilibrium phenomena, with holes ("antiparticles") with energy $\epsilon < 0$ included (we assume that the Weyl point is at $\epsilon = 0$).  In equilibrium, all our results reduce to the known  
chiral effects~\cite{DIMA,VOLOVIK,GRAVITY}.

\vskip0.2cm
{\bf 2.\,\,Berry curvature.}
The adiabatic motion of
a chiral quasi-particle close to the Fermi surface induces a Berry phase.
For completeness let us review the arguments presented in~\cite{ZAHED}
for representing this Berry phase as a WZW term. For a level crossing of degeneracy 2 in a Weyl
spectrum, the induced Berry phase $\vec{\cal A}(\vec{p})$  is a 
monopole of charge $g=1/2$ centered in the Fermi sphere. 

 The corresponding term in the action can be viewed as arising from  
an Abelian monopole in momentum space that encodes the effect of the level crossing and the U(1) anomaly. The Berry phase associated with a chiral quasi-particle is 
\be
{\bf S}_B=\int_{T_i}^{T_f}\,dt\,\vec{\cal A}(\vec{p})\cdot \dot{\vec{p}}
\label{B1}
\ee
where $\vec p(T_i)=\vec p(T_f)$. Since the Berry curvature is monopole-like and Abelian, the closed contour of the momentum $\vec p$ encircles the Dirac string and hence 
(\ref{B1}) is non-local. However it can be made local by extending it homotopically 
~\cite{WITTEN,CHIRAL}. For this purpose we introduce an auxiliary parameter $u\in[0,1]$ such that the momentum $\vec p(t)$ of a quasi-particle at the Fermi surface
 is extended as $\vec{p}(t)\rightarrow k_F\,\hat{p}(t,u)$,where $\hat{p}(t,0)=\hat{e}_3$ and  $\hat{p}(t,1)=\hat{v}$ is the particle velocity, see Fig. 1.
\begin{figure}
\includegraphics[width=8cm]{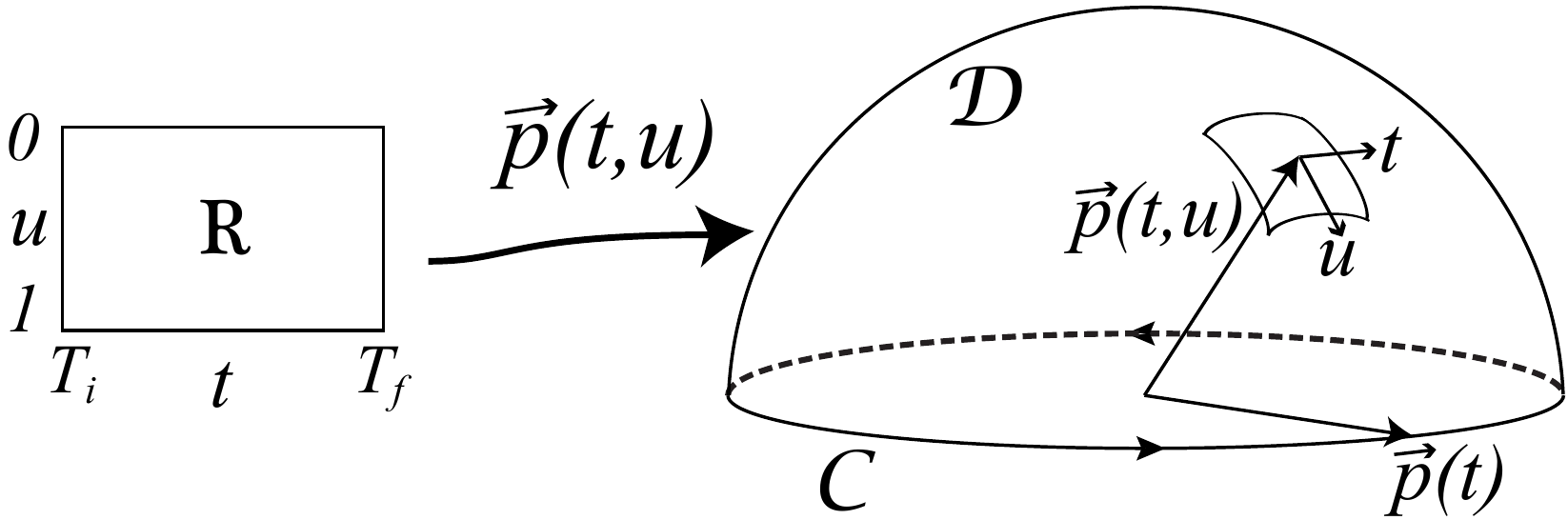}
\caption{For the momentum space monopole, the line integral (\ref{B1}) along the equator ($\cal C$) parameterized by $t$ can be re-written using the Stokes' theorem as a surface integral (\ref{B8}) over the upper hemisphere ($\cal D$) parameterized by $t, u$, where $u$ is an auxiliary variable. }
\end{figure}
The line integral (\ref{B1}) along $t$ can be written as a flux through the solid angle enclosed by $\vec p(t)$ parameterized by $(t,u)\in {\rm R} =[T_i,T_f]\times [0,1]$ by Stokes' theorem
\be
{\bf S}_B=\int_{\cal C=\partial {\cal D}}{\cal A} = \int_{\cal{D}}\,d{\cal A}
\label{B6}
\ee
where we use the form notation. The explicit form of the Berry potential in spherical coordinates is
$\vec{\cal A}(\vec{p})\cdot d\hat{p}= - g\left(1+{\rm cos}\,\theta\right)d\phi$,
with a Dirac string in the upper hemisphere. Therefore, the non-local form (\ref{B1}) can be written as the integral over the solid angle on the upper hemisphere, see Fig. 1: 
\be
{\bf S}_B
=\frac{g}{2k_F^3}\int_{\rm{R}}dt du\,
\epsilon^{ij}\epsilon^{abc} {p}^a\partial_i{p}^b\partial_j{p}^c .
\label{B8}
\ee
Eq. (\ref{B8}) is the WZW term for the Berry curvature of flux 1 or
monopole charge $g=1/2$ centered in a Fermi sphere of radius $k_F$; see \cite{VOLOVIK,HASAN,QI} for other uses of (\ref{B8}).

\vskip0.2cm
{\bf 3.\,\,Abelian chiral anomaly.}
To assess the effect of (\ref{B8}) on the transport of charged chiral fermions in  the vicinity of a Fermi surface 
at large $k_F$ and small temperature $T$,  let us introduce the coupling to electro-magnetism.  First, we formally
extend  $p^a\rightarrow p^\mu=k_F\ v^\mu(t,u,{\bf x})$, with $\mu=1,2,3,t,u$ and treat $v^\mu$ as a collective velocity on the
Fermi surface.  Let 

\be
dN_F=V_3\sum_{s=\pm 1} s\int d\Omega_k \,\left.{d^3k\over (2\pi)^3}\,{\bf f}(\epsilon,s)\right|_{k\approx k_F}
\label{000}
\ee
be the difference of particle ($s=+1$)  and anti-particle ($s=-1$) densities at the Fermi surface with a collective  velocity $v^\mu$ and energy $\epsilon$.  
For $k_F\gg T$, the coherent mean-field contribution of $dN_F$ to the anomalous 
effective action on the Fermi surface is
\be
d{\bf S}_F \equiv dN_F\,{\bf S}_B=\frac{g\ d\nu_F}{2k_F^3}
\int_{{\rm R}\times \mathbb R^3}\,p\,(dp)^2
\label{B10}
\ee
with  
\be
d\nu_F\equiv \frac{dN_F}{V_3}=n_F\sum_{s=\pm 1} d(s\epsilon)\,{\bf f}(\epsilon,s)
\label{B10X}
\ee
and 
$n_F=k_F^2/2\pi^2$. 
``Antiparticles" (holes with $\epsilon<0$) carry opposite charge and chirality and couple
to a monopole of opposite charge. Although they 
are exponentially suppressed at the Fermi surface, the antiparticles
contribute to the normalization
of the anomalous terms at finite temperature (see below).
Eq. (\ref{B10}) is odd under time-reversal; note that  
(\ref{B10X}) is even under $s\rightarrow -s$. In equilibrium
\be
{\bf f}(\epsilon,s)=\frac 1{e^{(\epsilon-s k_F)/T}+1}
\label{B10X1}
\ee

The full anomalous contribution follows from (\ref{B10}) 
after integrating $\epsilon$ across the Fermi sphere with $n_F$ fixed as the density of final states,
since all the new states generated by the anomaly show up at the Fermi surface due to Pauli blocking. In equilibrium, the energy integral is independent of temperature
\be
\left(\sum_{s=\pm}\int_0^\infty {d(s\epsilon)}\,{\bf f}(\epsilon,s)\right)\,=k_F .
\label{B14XX}
\ee
Gauging (\ref{B10}) by minimal substitution $p\rightarrow p+A$ with $F=dA$ ~\cite{NOTE}
thus leads to the anomalous effective  action 
\begin{eqnarray}
{\bf S}_F [A]=\frac{g}{4\pi^2}\int_{{\rm R}\times\mathbb R^3}\, (p+A)\left( dp+F\right)^2 .
\label{B11X}
\end{eqnarray}
The contribution $AF^2$ in (\ref{B11X}) is the 5-dimensional Chern-Simons term. Eq. (\ref{B11X}) 
is in agreement with the result in~\cite{ZAHED} derived at zero temperature.

The non-conservation of the 4-dimensional current follows from the gauge dependence of the action (\ref{B11X}) since under  $A\rightarrow A+d\theta$
\begin{align}
\delta_\theta{\bf S}_F [A]&=\frac{g}{4\pi^2}\int_{{\rm R}\times \mathbb R^3}\,\hskip-0.6cm d\theta\,(F+dp)^2 \nonumber\\
&= \frac{g}{4\pi^2}\int_{\mathbb R^4}\theta( F+dp)^2
 \equiv \int_{\mathbb R^4}d^4x\,\theta\,\partial_\mu{\bf J}^\mu_R .
\end{align}
In the second step we used the fact that the 4-dimensional space-time $\mathbb R^4$ is the boundary of the extended 5-dimensional space $ {\rm R} \times\mathbb R^3$ and Stokes' theorem. This derivation illustrates the anomaly inflow mechanism \cite{HARVEY}. Note that the 
contributions $dpF=d(pF)$ and $(dp)^2=d(pdp)$ can be absorbed into the constituent current. On the Fermi surface, they lead to the chiral magnetic and chiral vortical currents, i.e.
\be
\partial_\mu\left({\bf J}^\mu_R-\frac {gk_F}{4\pi^2}
 \epsilon^{\mu\nu\lambda\kappa}v_\nu F_{\lambda\kappa}-\frac{gk_F^2}{4\pi^2} \epsilon^{\mu\nu\lambda\kappa}v_\nu\partial_\lambda v_\kappa\right)=
 \frac{g}{2\pi^2}\,{\bf E}\cdot{\bf B}
\label{B14}
\ee
This result is expected to hold off-equilibrium provided that (\ref{B14XX}) is true on 
average. This is plausible if the applied external fields are time-independent.

\vskip0.2cm
{\bf 3.\,\,Non-Abelian chiral anomaly.}
To extend the preceding arguments to the non-Abelian case with $N$ flavors we assume for simplicity that
all Fermi momenta are the same, with a surface density 
$d\nu_F\rightarrow N d\nu_F$. The non-Abelian chiral anomaly
follows again by minimal substitution $p\rightarrow p+{\bf Q}^a{\bf A}^a$ with 
$N^2-1$ classical collective flavor charges ${\bf Q}^a$ which are the analogs of $\hbar \lambda^a/2$ in the limit $\hbar \rightarrow 0$ 
but ${\bf Q}^a$ fixed~\cite{DARBOUX}.  In equilibrium,  a rerun of the above arguments yields
 \begin{eqnarray}
&&\nabla_\mu
\left( {\bf J}^{a\mu}_R
-\frac {gN}{4\pi^2}\,\left<{\bf Q}^a{\bf Q}^b\right>\, k_F\,\epsilon^{\mu\nu\lambda\kappa}v_\nu\,{\bf F}_{\lambda\kappa}^b\right)\nonumber\\
&&=\frac{gN}{2\pi^2}\,\left<{\bf Q}^a{\bf Q}^b{\bf Q}^c\right>
{\bf E}^b\cdot{\bf B}^c
\label{ZB14X}
\end{eqnarray}
The color averaging for $SU(N)$ is carried using $N-1$ delta functions ensuring the conservation of the
$N-1$ Casimirs. Specifically $\left<{\bf Q}^a{\bf Q}^b\right>=\delta^{ab}/2$ and 
$\left<{\bf Q}^a{\bf Q}^b{\bf Q}^c\right>=d^{abc}/4$,
in agreement with~\cite{STONE,SON1,MISHA,SSZ,ZAHED} for charges with equal
Fermi momenta. 
Again (\ref{ZB14X}) is expected to hold
off-equilibrium provided that (\ref{B14XX}) holds on average.

\vskip0.2cm
{\bf 4.\,\, Rotating frames.} 
Let us now present an alternative view on the CVE. 
Consider a frame rotating around ${\bf\hat{z}}$ in the presence of a weak gravitational potential $\varphi$. A rotating frame can be described by 
a stationary metric with $g_{00}= 1-\Omega^2\rho^2+2{\varphi}$ and
$g_{0\phi}= -\Omega \rho$ where $\rho=x^2+y^2$.
To linear order, the motion of a quasi-particle in such a background can be described via an energy dependent gauge field 
$\hat{A}^a_g$ which can be constructed with the help of the veirbeins using $g_{\mu\nu}=e^a_\mu e^b_\nu\, \eta_{ab}$. 
Indeed, to linear order $e^a_\mu\equiv\delta^a_\mu+\hat e^a_\mu+\dots$ and
\begin{eqnarray}
g_{\mu\nu}\,p^\mu p^\nu&\approx&(p^a+\hat e^a_\mu\, p^\mu) (p^b+\hat e^b_\nu\, p^\nu)\,\eta_{ab}\nn
&\equiv&(p^a+  A_g^a) (p^b+ A_g^b)\,\eta_{ab} ;
\end{eqnarray}
the corresponding gravi-electro-magnetic fields are ${\bf E}_g=p^0 \nabla \varphi$ and
${\bf B}_g=p^0\,{\bf \Omega}$. 
The inertial 3-force on a quasi-particle of 4-momentum 
$p^a\equiv(E, {\bf p})$ is simply the corresponding Lorentz force
\be
{\bf F}=E\,\left( \nabla \varphi+ {\bf v}\times s\,\bf{\Omega}\right)
\label{RF2}
\ee
and the second term in (\ref{RF2}) is the relativistic Coriolis force. The factor $s$ arises because under time-reversal ${\bf \Omega}\rightarrow -{\bf \Omega}$. 

The effect of the gravi-electro-magnetic fields on 
the Fermi surface with $(E, {\bf p})\rightarrow (\epsilon,s{\bf k}_F)$ will be of the form ${\bf E}_g\cdot{\bf B}_g=\epsilon^2 \nabla\phi\cdot\Omega$. 
The corresponding Chern-Simons term is
\be
\frac{g}{4\pi^2}\left(\sum_{s=\pm}\int_0^\infty \frac {d(s\epsilon)}{\epsilon}\,{s\,\epsilon^2}\,{\bf f}(\epsilon,s)\right)\,
\int_{{\rm R}\times \mathbb R^3}\, A\, {\bf F}_G^2
\label{MAN}
\ee
with ${\bf F}_G\equiv {\bf F}_g/p^0$. 
In contrast to the electro-magnetic case above, the density of states in (\ref{MAN}) is not kept fixed at the value $n_F$, thus 
$\epsilon$ rather than $k_F$ appears in the denominator of (\ref{MAN}). 
The rotation induces an energy dependent force (\ref{RF2}) which affects every mode in the Fermi sphere proportional to its energy $\epsilon$. As a result, the anomaly flow moves all modes within the Fermi sphere and Pauli blocking does not apply. In equilibrium
\be
\left(\sum_{s=\pm}\int_0^\infty \,d\epsilon\,\epsilon\,{\bf f}(\epsilon,s)\right)\,=\frac 12\left(k_F^2+\frac{(\pi T)^2}3\right)
\label{RF5}
\ee
Off equilibrium (\ref{RF5}) can be modified; however the $k_F^2$ term is protected by topology and gauge symmetry
\cite{MISHA,SSZ,ZAHED}.

The anomalous contribution to the current stemming from (\ref{MAN}) therefore is
\be
\partial_\mu\left(
{\bf J}^\mu_R-\frac{g}{4\pi^2}\,(C+\varphi)\,\left(k_F^2+\frac{(\pi T)^2}3\right)
 \,(0,{\bf \Omega})^\mu\right)=0
\label{RF6}
\ee
The constant $C=2$ is fixed by the $(vdv)^\mu=(0, 2\Omega{\bf z})^\mu$ term in 
(\ref{B14}); see also ~\cite{MISHA,SSZ,ZAHED,VILENKIN,VOLOVIK,GRAVITY}.

The new element in our
derivation is the appearance of $\varphi$ which we will now exploit. A small and in-homegeneous temperature variation 
along the rotating direction, say $\Delta T=T(z_1)-T(z_0)>0$,  on top of the Fermi sphere is captured by a gravitational potential $\varphi(z)$ such that $\varphi(z_0)-\varphi(z_1)=\Delta T/T$. This follows from an observation by Luttinger~\cite{LUTTINGER}
that the effect of a temperature gradient driving the system out of equilibrium can be balanced by a gravitational potential. 
Thus, the response of the system to a temperature gradient can be captured by the response to a gravitational potential, as is
clear in (\ref{RF6}) through $(1+\varphi/2)\,{(\pi T)^2}/3 \approx {(\pi(T+\Delta T/4))^2}/3$. A temperature gradient along the axis of rotation thus leads to a novel contribution to the chiral vortical current:
\be
{\bf J}_R = \frac{g}{4\pi^2}\,\frac{\Delta T}{T}\,\left(k_F^2+\frac{(\pi T)^2}3\right)
 \, {\bf{\Omega}} .
 \ee

\vskip0.2cm
{\bf 5.\,\, Anomalous chiral superfluids.}
The non-universality of the temperature induced effects in the CVE can also be seen by considering
chiral superfluids~\cite{MISHA,SSZ,NAIR}. In the leading mean-field approximation, the right
constitutive flavor currents in a rotating chiral fluid are~\cite{MISHA} ($a=1,...,N^2-1$)
\be
{\bf J}^{ia}_R\approx F_\pi^2\mu\,e^a{\bf v}^i+\frac {{\cal C}}2\,d^{abc}e^be^c\,{\bf \Omega}^i \left(\mu^2-\frac{3N}{8}
\frac{\mu^2}{F_\pi^2}\,
\sum_k\frac {n_{k}}{\omega_k}\right)
\label{TAD1}
\ee
with $\mu^a/\mu=e^a$ and  ${\cal C}$ the non-Abelian anomaly normalization. 
The first contribution is the normal fluid contribution, while the second contribution is
the anomalous contribution due to the SU(N) Goldstone modes in the  superfluid phase, 
with $F_\pi$ their weak decay constant. 

A specific  realization of the chiral
superfluid is captured by the effective approach to the color-flavor locked (CFL) phase of dense 
QCD with $N=N_f=N_c$~\cite{CFL}. 
For the CFL superfluid $n=2N^2(\mu^3/6\pi^2)$ and $F_\pi^2=n/\mu$.
At small but finite temperature, $n_k$ is the Bose distribution for 
the Goldstone modes as pion-diquarks, so that
\be
\sum_k\frac {n_{k}}{\omega_k}\equiv \int_0^\infty
\frac{kdk}{2\pi^2}\,\frac 1{e^{k/T}-1}=\frac{T^2}{12}
\ee
Inserting these results in the CVE current (\ref{TAD1}) yields
\be
{\bf J}^{ia}_R\approx n\,e^a{\bf v}^i+\frac {{\cal C}}2 d^{abc}e^be^c\,{\bf \Omega}^i \left(\mu^2-\frac 38\,
\frac{{(\pi T)^2}}{12}\right)
\label{TAD2}
\ee
which is to be compared to  (\ref{RF6}). Eq. (\ref{TAD2}) is the right 
SU(N) flavor current in the superfluid CFL phase of dense QCD with (pion-diquark) 
bosonic  fluctuations on a fully gapped Fermi surface whereas (\ref{RF6}) is the right U(1) current 
on a chiral Fermi surface threaded by a Berry phase with fermionic fluctuations.  
 The temperature corrections to the CVE in (\ref{TAD2}) and (\ref{RF6}) 
differ in both sign and magnitude. In the chiral superfluid
higher order temperature corrections are readily available beyond the leading mean-field
approximation in the form $\mu^2((\pi T)^2/F_\pi^2)^n\approx ((\pi T)^2/\mu^2)^n$.

\vskip0.2cm
{\bf 6.\,\,Mixed gravitational anomaly.}
Let us now discuss the connection of our CS-based approach to the chiral gravitational anomaly. 
Chiral fermions on a curved manifold couple to the spin connection, 
leading to a mixed U(1) and gravitational anomaly. In our semi-classical analysis this
is achieved through minimal substitution
$p_\mu\rightarrow \left(p_\mu+A_\mu\right){\bf 1}_4+{\bf S}_{ab}{\bf \omega}_\mu^{ab}/4$,
where ${\bf \omega}_\mu^{ab}$ is the anti-symmetric spin connection
and ${\bf S}_{ab}$ is the antisymmetric spin matrix. Thus, the mixed gauge-gravity CS action
at the Fermi surface is
\begin{eqnarray}
\frac {g}{4\pi^2}\int_{{\rm R}\times \mathbb R^3}
 \left<A {\bf 1}_4+\frac 14 {\bf S}_{ab}{\bf \omega}^{ab}\right>
\left<\left(F{\bf 1}_4 +\frac 14 {\bf S}_{cd}{\bf R}^{cd}\right)^2\right>\nonumber\\
\label{Z2}
\end{eqnarray}
with ${\bf R}=d{\bf \omega}+{\bf \omega}^2$ the full Riemann tensor thanks
to Lorentz gauge-invariance. The averaging includes the tracing and  is carried over the 
spin measure with pertinent constraints for the Casimirs. Specifically, $\left<{\bf 1}_4\right>=1$ and
$\left<{\bf S}_{ab}{\bf S}_{cd}\right>=({4\,{\bf s}^2}/{3})\left(\eta_{ac}\eta_{bd}-\eta_{ad}\eta_{bc}\right)$,
with ${\bf s}^2$ the squared spin vector. A rerun of the
above arguments for the U(1) right current gives the mixed covariant anomaly
\be
\frac 1{\sqrt{-{\bf g}}}\partial_\mu
\left(\sqrt{-{\bf g}}\,{{\bf J}}^\mu_R\right)=\frac{g}{8\pi^2}\,\left(F^{\mu\nu}\tilde{F}_{\mu\nu}-\frac {{\bf s}^2}{6}\,{\bf R}^{\alpha\beta\mu\nu}{\tilde{\bf R}}_{\alpha\beta\mu\nu}\right)
\label{Z3}
\ee
with the subsumed anomalous contributions in ${{\bf J}}_R$. In our semi-classical approach, the squared spin length is given by ${\bf s}^2=(1/2)^2$ as 
different spin components commute when $\hbar \to 0$. For this semi-classical ${\bf s}^2$ value 
we thus reproduce the correct coefficient of the mixed gravitational anomaly 
(note
that the quantum value ${\bf s}^2=3/4$ would result
in a coefficient that is larger by 3).

\vskip0.2cm
{\bf 7.\,\,Conclusions.}
The minimally gauged WZW term on a Fermi surface 
with a Berry phase is at the origin of most anomalous contributions to the 
chiral currents.
It embodies aspects of the Callan-Harvey anomaly in-flow~\cite{HARVEY},
as per the descent equation in cohomology~\cite{ZUMINO}.
Although our approach is not holographic, the induced 5-dimensional Chern-Simons term
acts like its holographic analog.

The matter effects do not modify the divergence of the chiral current given by the anomaly,
but generate additional gauge-invariant anomalous contributions to the currents themselves. In particular, for rotating systems with chiral Fermi surfaces, we observe a new contribution to the CVE resulting from a temperature gradient.

Our formalism based on the 5D CS action can be applied to inhomogeneous systems
in- and off-equilibrium. 
The temperature corrections to the CVE are different for chiral Fermi surfaces and 
chiral superfluids. However in both cases the leading anomalous terms proportional to $k_F^2$ are not renormalized due   
to gauge symmetry and topology.

Recently, an extended kinetic formulation with
a Berry monopole was suggested in~\cite{SON,STEPHANOV,Chen:2012ca,STONE} where the anomaly is sourced at the center 
of the Fermi sphere, whereas in our treatment we focus on the properties of the Fermi surface. Nevertheless, these approaches are related since the incompressibility of the Fermi sphere forces the fermions produced by the anomaly to appear on the Fermi surface.   

\vskip0.2cm
{\bf Acknowledgements.}
We thank Hans Hansson, Misha Lublinsky, Michael Stone, Grigory Volovik,  and Ho-Ung Yee for discussions.
  This work was supported by the U.S. Department of Energy under Contracts No.
DE-FG-88ER40388 and DE-AC02-98CH10886.


\end{document}